\documentclass[11pt]{article}

\usepackage[dvips]{graphicx}
\DeclareGraphicsExtensions{eps,.eps,.ps}
\graphicspath{{./}}
\begin{document}

\baselineskip=14.20pt

\begin{center}
{\bf Characterization of the Micromechanics in a Compressed Emulsion
System: Force Distributions.}
 
{ Jasna Bruji\'c$^1$, Sam F. Edwards$^1$, Ian
Hopkinson$^2$, and Hern\'an A. Makse$^3$}

{   
$^1$ Polymers and Colloids Group, Cavendish Laboratory, University of
Cambridge, Madingley Road, Cambridge CB3 0HE, UK\\
$^2$ Department of Physics, UMIST, Sackville Street, Manchester
M60 1QD, UK\\
$^3$ Levich Institute and Physics Department, City College of New York, 
New York, NY 10031, US 
}

\end{center} 

\begin{abstract}
The micromechanics of a variety of systems experiencing a structural
arrest due to their high density could be unified by a thermodynamic
framework governing their approach to 'jammed' configurations. The
mechanism of supporting an applied stress through the microstructure
of these highly packed materials is important in inferring the
features responsible for the inhomogeneous stress transmission and
testing the universality for all jammed matter.  In this paper we
present a novel method for measuring the force distribution within the
bulk of a compressed emulsion system using confocal microscopy and
explain our results with a simple theoretical model and computer
simulations.  We obtain an exponential distribution at large forces
and a small peak at small forces, in agreement with previous
experimental and simulation data for other particulate systems.

\end{abstract}

\begin{center}
(submitted to Proc. Nat. Acad. Sci.)
\end{center}
\newpage

\section{Introduction}
\label{intro}

The concept of jamming is emerging as a fundamental feature of many
systems with slow relaxation dynamics such as granular matter, complex
fluids and structural glasses \cite{Jamming,sam-zero,Cipelletti2001}.
 Whereas one can think of liquids or suspensions as consisting of
particles which move very slowly compared to gases, there comes a
point where all particles are in close contact with one another and
therefore experience structural arrest.
In granular systems and
compressed emulsions there is no kinetic energy of consequence;
the typical energy required to change the positions of the
jammed particles is very large compared to the thermal energy at room
temperature. As a result, the material remains arrested in a
metastable state and is able to withstand an applied
stress \cite{Cates1999}.

There is a growing literature studying the ''jammed'' state in
particulate assemblies, aiming to characterise its
micromechanics \cite{Jamming}. It has been shown experimentally that
the stresses are distributed inhomogeneously through granular
materials and the features of the distribution are useful in inferring
the structural elements associated with mechanisms of supporting the
applied stress.
In order to develop a theory to describe such closely packed
particles one needs to know the geometry of the packing
in the bulk, and the distribution and propagation of stress
in these systems.

Several approaches have so far been employed,
including 2D and 3D experimentation \cite{dantu,Liu1995}, numerical
simulations \cite{Radjai1996,Thornton,Makse2000,Antony2001,Ohern00}
and statistical
modelling \cite{Coppersmith95}.  Previous experiments in 3D
assemblies have been confined to measurements of the probability
distribution of forces exerted at the boundaries with the container,
thus reducing the dimensionality of the problem \cite{Liu1995,
Mueth1998,Lovoll1999,Blair2001,Makse2000}.
These measurements provide a
quantitative understanding of the inhomogeneity of stress transmission
within the bulk. However, the method does not have access to the
spatial arrangement of the contact force network and other
structural features,
such as force chains and arching, which have been postulated as the
signature of jamming \cite{Jamming,Ohern00}.

The salient feature of the probability distribution $P(f)$
of interparticle contact force $f$ in jammed systems,
obtained from the above methods,
is
an exponential decay above the mean value of
the force. This feature of $P(f)$  seems very robust,
with growing evidence that it is independent
of particle rigidity \cite{Radjai1996,Antony2001},
crystallinity \cite{Blair2001}, tangential forces, construction history and
friction
\cite{Makse2000}. Nevertheless, there is no clear consensus on the general
functional form
of $P(f)$ as there are significant discrepancies in the literature
particularly regarding the behaviour at small forces, both between
experimental data and the theoretical model predictions. Moreover,
the
possibility of a crossover to a  Gaussian-like
distribution has been observed at large confining pressures
\cite{Thornton,Makse2000,Sexton1999}.

In this paper we present a novel method to measure the force
distribution within a concentrated emulsion system in its jammed state,
which provides the complete three-dimensional information of the contact
force network and the spatial arrangement of droplets.  We address this 
problem using confocal microscopy, which provides direct measurement of the
dispersed phase morphology within the bulk of the sample. The emulsion
droplets are compressed by an external pressure through centrifugation
because of the density difference between the phases, and a force network
develops within the system in response to the applied stress. At high
volume fractions, above the random close packing regime, emulsions
exhibit an elasticity which is rationalised by the storage of energy
through the deformation of droplets, given in terms of their Laplace
pressure \cite{Princen1983}. The degree of deformation is used to
derive an interdroplet force. The 3D imaging of a whole ensemble allows
the calculation of the forces between the droplets, thus enabling the
determination of $P(f)$. We find that the distribution is 
characterised by a small peak at low forces and an exponential decay at forces above the mean value, a result that can be described by the functional form of $P(f)$ derived from the simple theoretical model we propose in this paper.

The form of the probability distribution is
independent of the material of the particle provided it has well
defined elastic properties. Therefore we can expect the
micromechanics of an emulsion, comprised of very ``soft'' particles, to bear
many similarities with a packing of granular materials, such
as ball bearings or glass beads.
Even though there have been no studies
of $P(f)$ specifically devised for compressed emulsion systems, we
compare our results with the existing data for other jammed systems
such as grains and foams, thus testing the hypothesis of a common
behaviour for all such matter. 

We use numerical simulations to examine the effect of polydispersity, osmotic pressure, and
other microstructural features, such as the coordination number,
on the distribution of forces. They are designed to mimic the experimental procedure on
monodisperse and polydisperse distributions of soft particles. The
numerical simulation result at the appropriate confining pressure is in good agreement with that
 obtained from the experiment, showing that the form of the force distribution is indeed independent of particle polydispersity.

Moreover, we offer in this paper
what we believe is the simplest realistic theory of the force
distribution, a theory which does not attempt the ambitious study of
the percolation of forces \cite{Bouchaud}, but has the advantage of a
simple analytic solution. The theory is sufficiently crude that the
reader will be able to see all sorts of ways in which it can be
improved, however the simple prediction of the theory can easily be
compared to experimental results and is found to describe the data well.

\section{Experiment}
\label{exper}       

We use a Zeiss LSM510 confocal laser scanning microscope equipped with
a high numerical aperture oil-immersion objective lens with a 40$\times$
magnification. The fluorescent dye is excited with a
488nm Argon laser and the emitted light is detected using a photomultiplier behind a long-pass 505nm filter. These settings are appropriate for the excitation of Nile Red dye, used to label the emulsion described below [see 
Ref. \cite{faraday} for more details]. 
The sample volume (76.3$\times$76.3$\times$23.6$\mu$m) is
typically acquired from regions 30$\mu$m below the upper surface of
the sample. In this work a 3D image is acquired in approximately 2
minutes.

The emulsion system constitutes of silicone oil droplets in a
refractive index matching solution of water ($w_t=50\%$) and glycerol
($w_t=50\%$), stabilised by 0.01$mM$ sodium dodecylsulphate
(SDS). This system is a modification of the emulsion reported by Mason
{\it et al.}  \cite{Mason1995} to produce a transparent sample
suitable for confocal microscopy.  

 The droplet size distribution, measured by image analysis,
gives a mean radius of 3.4$ \mu m$ with a radius range between 1$
\mu m$ and 10$ \mu m$.  This relatively
narrow droplet size distribution is achieved by applying high shear
rates ($7000s^{-1}$) to a coarsely mixed parent emulsion using a
Linkam shear stage \cite{Perrin2000}. 
To provide contrast between the phases in
the microscope, the dispersed oil phase is fluorescently labelled by
adding 0.1mM solution of Nile Red dye, predissolved in acetone. 
The
emulsion system prepared in this way remains stable to coalescence for
at least a year.

\begin{figure}
\centering
{\resizebox{10cm}{!}{\includegraphics{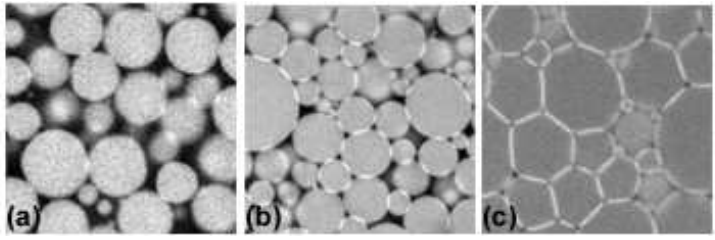}}}
\caption{2D slices of emulsions under varying compression rates: 
1g(a), 6000g(b) and 8000g(c).}
\label{slices}
\end{figure}

The threshold volume fraction for the onset of elasticity depends on
the polydispersity of the emulsion, or in other words, the efficiency
of the packing. The sequence of images in Fig. \ref{slices} shows 2D
slices from the middle of the sample volume after: (a) creaming under gravity, (b) centrifugation at 6000g for 20 minutes and (c) centrifugation at 8000g for 20 minutes. The samples were left to equilibrate for several hours prior to measurements being taken. 
The
volume fraction at the onset of droplet deformation 
for our polydisperse system  is $\phi= 0.90$, 
determined by image analysis. 
This high volume fraction obtained at a relatively small 
osmotic pressure of 125 Pa is achieved due to the polydispersity of
the sample.

Confocal imaging of the static sample revealed an 
effect which occurs upon emulsion compression. The areas of contact
between the droplets fluoresce with a higher intensity than the
undeformed perimeters on the bodies of the droplets, thus highlighting
the regions of interest. Images presented in Fig. \ref{slices} 
illustrate this trend 
as the 
osmotic pressure is increased. This effect can be attributed to the
increase in dye concentration at the regions of deformation as
two droplet surfaces are pushed together to distances smaller than the 
resolution of the microscope. 
The fluorescent dye has an
affinity for the surfactant so that an enhanced surfactant
concentration leads to an enhanced dye concentration and thus higher
fluorescence. 
Future work will involve a more thorough investigation
of this effect.

\subsection{The Force Model}
\label{forcemodel}

The forces between the droplets are
calculated from the 3D images (of size 256$\times$256$\times$64 voxels)
by means of existing interdroplet force models.
We extract the positions and
radii of all the droplets with subvoxel accuracy using a Fourier Filtering
Method (FFM) \cite{Parker,faraday}.
The areas of contact patches shown in Fig. \ref{slices}(b)
are extracted based on an intensity threshold, since they are brighter
than either the droplets or the aqueous background. 

The determination
of an accurate force model for the compression of two droplets is not
trivial, but can be simplified within certain limits. For small
deformations with respect to the droplet surface area, the Laplace
pressure remains unchanged and all the energy of the applied stress is
presumed to be stored in the deformation of the surface. Hence, at the
microscopic level, two spherical droplets in contact with radii $R_1$
and $R_2$ interact with a normal force
\begin{equation}
\label{force}
f = \frac{\sigma}{\tilde{R}} ~A
\end{equation}                                                                 This is the Princen model \cite{Princen1983}, where $A$ is the area of deformation, $\sigma$ is the interfacial tension of the droplets and
$\tilde{R}$ is the geometric mean of the radii of the undeformed
droplets, $\tilde{R}=2 R_1 R_2/(R_1+R_2)$. The normal force acts
only in compression, i.e. $f = 0$ when there is no overlap.

The above force corresponds to an energy of deformation which is
quadratic in the area of deformation, analogous to a harmonic
oscillator potential that describes a spring satisfying Hooke's
law. There have been several more detailed
calculations \cite{Witten1993} and numerical
simulations \cite{Lacasse1995} to improve on this model and allow for
anharmonicity in the droplet response by also taking into
consideration the number of contacts by which the droplet is
confined. Typically these improved models lead to a force law for
small deformations of the form $f \propto A^\alpha$ , where $\alpha$
is a coordination number dependent exponent in the range 1-1.5. 
Nevertheless, this work assumes the Princen model, the validity of which was tested by the summation of all the forces on a single droplet. Since the sum on each droplet could be approximated to zero we consider the method adequate.

\subsection{Experimental results}
\label{exp-results}

\begin{figure}
\centering
{\resizebox{10cm}{!}{\includegraphics{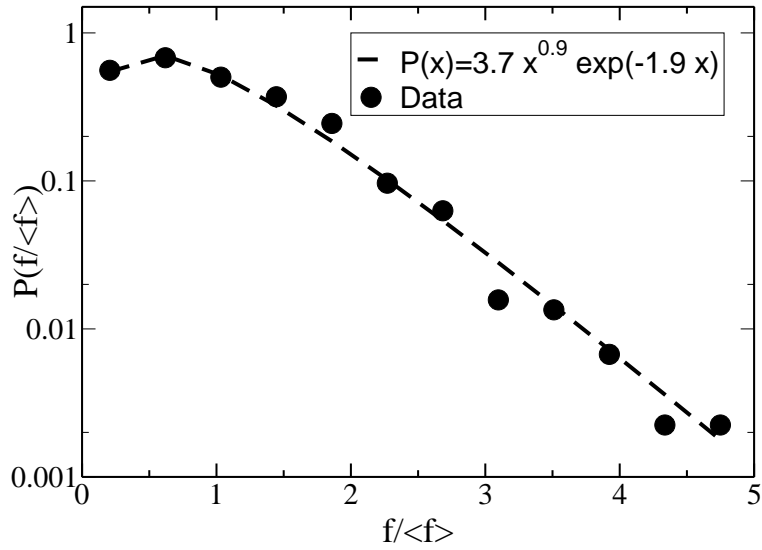}}}
\caption{Probability distribution of the contact forces for the
compressed emulsion system shown in Fig. \protect\ref{slices}(b).
We also show a fit to the theory developed in Section \protect\ref{theory}.}
\label{pf}
  \end{figure}

Figure \ref{pf} shows the probability distribution of interdroplet
forces, $P(f)$, for the sample shown in Fig. \ref{slices}(b). We use
the Princen model (Eq. (\ref{force}), $\sigma = 9.8 \times 10^{-3}$N/m
\cite{Mason1995}) to obtain the interdroplet forces from the contact
area data extracted from the image analysis described above.

The forces are calculated from the bright, fluorescent patches that
highlight the contact areas between droplets.
The radii of the droplets needed to obtain the forces
according to Eq. (\ref{force}) are obtained with the FFM. 
The distribution data shown are extracted from
1234 forces arising from 450 droplets. The data shows an exponential
distribution at large forces, consistent with results of many
previous experimental and simulation data on granular matter, foams,
and glasses. The behaviour in the low force regime indicates a small
peak, although the power law decay tending towards zero is not well
pronounced.  The best fit to the data gives a functional form
of the distribution $P(f)\propto {f}^{0.9}e^{-1.9f/\bar{f}}$,
consistent with the theoretical model proposed in Section \ref{theory}
and the existing literature \cite{Coppersmith95}. It is inappropriate
to draw conclusions on the physical significance of these
coefficients, since the geometry of the packing in the experiment is
very different to our simple theoretical model.

Our experimental data allows us to examine the spatial distribution of the forces in the compressed emulsion, shown in Fig. \ref{chains}.
In this admittedly small sample volume, the forces appear to be uniformly distributed in space and do not show evidence of localisation of forces
within the structure. Moreover, we find that the 
average stress 
is independent of direction, indicating isotropy.  
Other experiments are underway to probe the existence of force chains in compressed emulsion systems. 

\section{Theory}
\label{theory}

Although the experimental system consists of polydisperse particles
which are deformable, in order to get a tractable theory, we simply
consider spheres in multiple contact greater than or equal to 
four in 3-D. Even though this is a gross simplification we
believe that a theory that can be carried through to an analytic
solution is worthwhile. The reader is referred to more comprehensive
theories such as the q-model \cite{Coppersmith95} or force-splitting
models \cite{Bouchaud} for more detailed analysis, which are
correspondingly more difficult to solve.

By Newton's laws, in equilibrium the sum of all the forces exerted on a
particle by its nearest neighbours is zero.  In three dimensions the
average shape of a particle is a sphere, and the minimum  co-ordination
number $N$ is four \cite{Z}.
Consequently the force
$\vec{f}$ exerted by a particle on one of its neighbours will equal
the sum of the forces $\vec{f}_1+\vec{f}_2+\vec{f}_3$ 
of the other neighbours in
contact with it. To simplify we consider the scalar $f=|\vec{f}|$
since it will have a very similar distribution on every grain. The
distribution of the vector $\vec{f}$ will differ even on adjacent
grains, therefore we calculate $P(f)$, not $P(\vec{f})$. It is
important to note that only those forces which are pushing on each
particle are taken into account in the calculation of $P(f)$, a fact which will appear in the range of integration. 

Our model takes into account the direction cosines of
each of the forces. 
Excluded
volume is also an important factor as particles  cannot overlap
and the four particle force correlation function should be
included. All these effects can be crudely modelled by blurring the
contribution from each of the pushing forces 
by a factor $\lambda_i \in[0,1] (i=1, 2, 3)$,
which plays the role of
the direction cosine and the other correlation factors.
 
A force balance equation which is capable of analytic solution is
\begin{equation}
        f=\lambda_1^2 f_1 +\lambda_2^2 f_2+\lambda_3^2 f_3.
     \end{equation}
This gives rise to an equation of the  Boltzmann form:

\begin{equation}
        P(f)=\int_0^{\infty} df_1 df_2 df_3
\int_0^1
d\lambda_1 ~d\lambda_2 ~d\lambda_3~\delta(f-\lambda_1^2 f_1-\lambda_2^2 
f_2-\lambda_3^2 f_3) P(f_1)P(f_2) P(f_3)
\end{equation}

It is convenient to work with the Fourier transform of the probability
distribution ${\cal{P}}(k)=\int_{-\infty}^{\infty} e^{ikf} P(f) df$
which gives
\begin{equation}
        {\cal{P}}(k)=(\int_0^1 d\lambda{\cal{P}}( \lambda^2 k))^3,
\end{equation}
which can be solved to give the normalised distribution
\begin{equation}
        P(f)=\frac{2}{\sqrt{\pi}} \frac{f^{1/2}}{p^{3/2}} e^{-f/p},
\end{equation}
where $p \propto \bar{f}$ and the
proportionality constant depends on the exponent of the power law rise
at low forces.  More generally, if there are $N$ contacts arising from
differing geometric configurations, similar calculations give 
\begin{equation}
        P(f) \propto f^{1/(N-2)} e^{-f/p}.
\end{equation}
 
Note that for a large number of contacts, $P(f)$ reaches zero very near $f=0$. 
In Fig. \ref{pf}
we see a comparison between the theoretical form and the experimental
data in good agreement. 
There are many improvements which are essential for belief in
coefficients, but the functional form, starting at zero and ending
with an exponential decay, seems well founded.

\section{Simulations}

We perform molecular dynamics (MD) simulations to gain insight into 
the effects of osmotic pressure, polydispersity, and other microstructural 
features such as the coordination number and force chains on the
probability distribution, $P(f)$. 
The numerical protocol is designed to mimic the experimental 
procedure used to prepare compressed emulsion systems at different 
osmotic pressures, described in Section \ref{exper}.
Our model considers an assembly of deformable spherical droplets
interacting via repulsive normal forces given by 
the Princen model in Eq. (\ref{force}).
The continuous liquid phase is modeled in its simplest form, 
as a viscous drag
force acting on every droplet, proportional to its velocity.
The dynamical evolution of the droplets is obtained  by solving Newton's
equation for an assembly constrained by a given osmotic pressure.
Our model is
similar to the Discrete Element Method (DEM)
\cite{cundall,Makse2000}
 used in
MD simulations of granular materials. However, we
adjust the DEM for the system of compressed emulsions by
exclusion of transversal forces (tangential elasticity and 
Coulomb friction)
and by computation of interparticle
forces using the principles of interfacial mechanics described by the
Princen model instead of the Hertz model, often used in contact mechanics
of solid particles $(f\sim \xi^{3/2}$).

The simulations begin with a set of non-overlapping
2000 spherical particles located at
random positions in a periodically repeated cubic cell of side
$L$. 
At the outset, a
series of strain-controlled isotropic compressions and expansions are
applied until a volume fraction slightly below the critical density of
jamming is reached
\cite{Makse2000}.   The system is then compressed and extended slowly until a
specified value of the stress and volume fraction is achieved at
static equilibrium. The distribution of forces within the static
structure is calculated and
then directly compared to that obtained from
experiments and theory.

 We first consider a quasi-monodisperse system composed of 1000 droplets
of radius 1.05 $\mu$m and 1000 droplets of radius $0.95\mu$m.
 Then the  effect of polydispersity is investigated by
consideration of the radii distribution obtained from our
experiments  characterized by a Gaussian distribution
with a mean value
 $<R>=3.4\mu$m and standard deviation 1.44$\mu$m, and a distribution
range between 1$\mu$m and $6.6\mu$m.
The osmotic pressure, $\Pi$,  
is varied between 1 Pa and 1 kPa,
again mimicking the experiments.

\begin{figure}
\centering
{\resizebox{10cm}{!}{\includegraphics{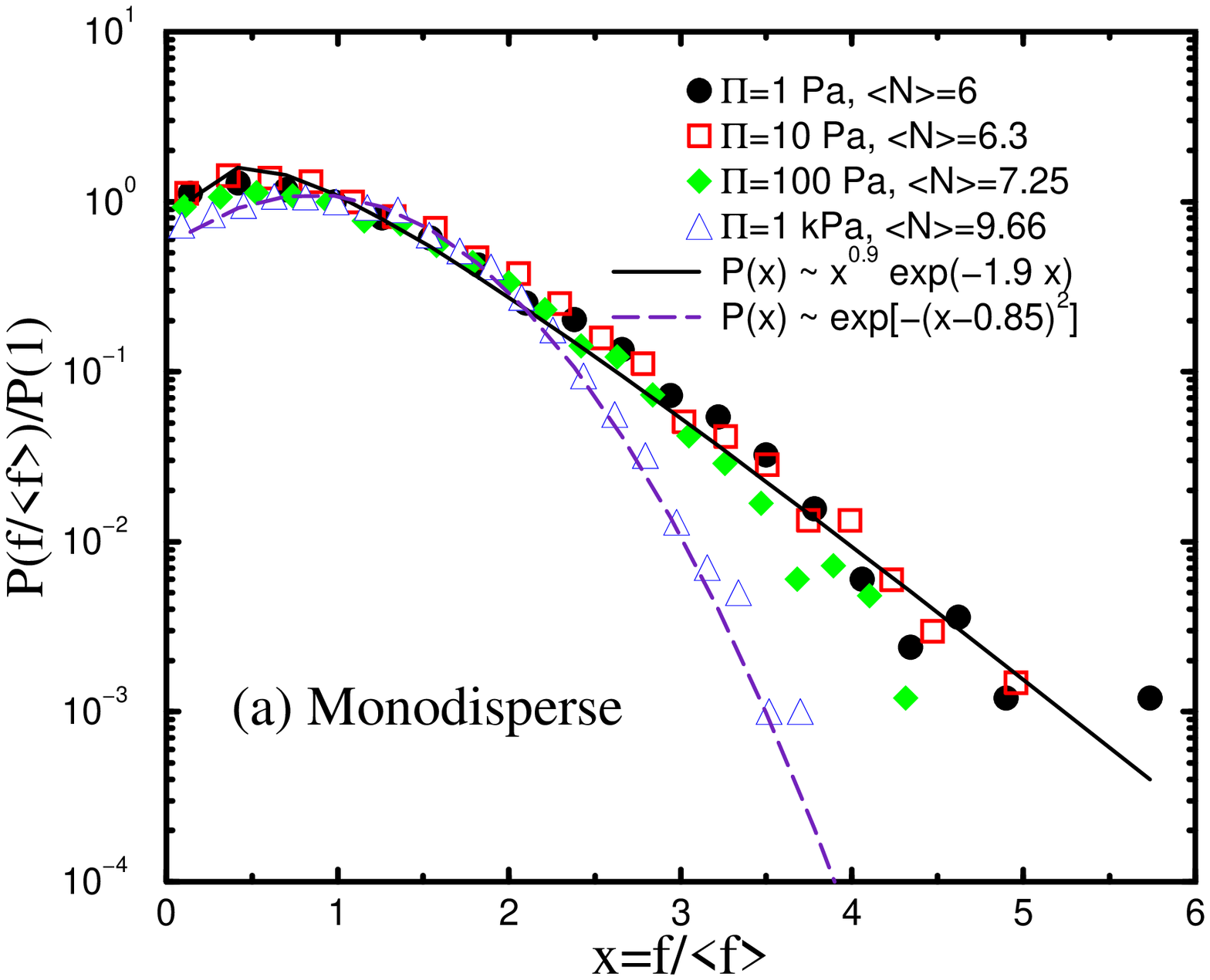}}}
\centering
{\resizebox{10cm}{!}{\includegraphics{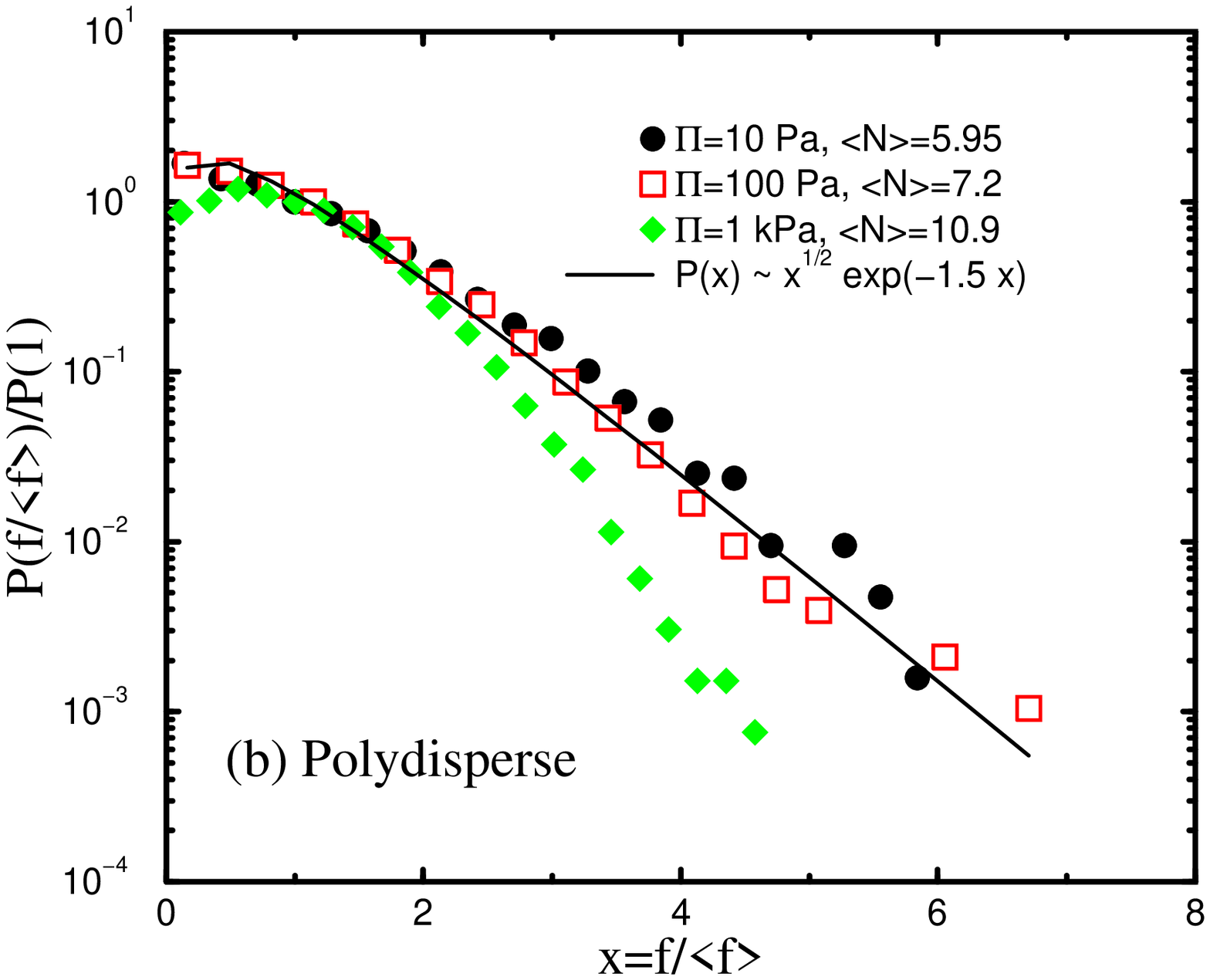}}}
\caption{Numerical results for $P(f)$ for a system
of (a) monodisperse and (b) polydisperse emulsions at different
osmotic pressure, $\Pi$ , and mean coordination number $<N>$.
}
\label{simulations}
\end{figure}

Figure \ref{simulations} shows the results of the simulations.  We
see that the simulated data for monodisperse and polydisperse
systems at low osmotic pressure agrees with the experiments
and the theory. At low pressures the system is close to jamming (near
RCP at $\phi \sim 0.64$) and the average coordination number is close
to its minimal value $<N>=6$ for particles interacting by normal
forces only as given by constraint arguments \cite{Z,Makse2000}.  At
large pressures, when the coordination number significantly departs
from its minimal value, the probability distribution departs from the
prediction of the theory and crosses-over to a Gaussian-like
distribution in the case of the monodisperse system. In the case of
the polydisperse system, the distribution at large pressures departs
from the exponential decay at large forces, but its form cannot be
fitted by a Gaussian-like distribution.

The numerical simulation performed under the same conditions as in the
experiment yields a $P(f)$ of the same functional form for the
appropriate osmotic pressure ($\Pi\sim 100$ Pa) and polydispersity
(Fig. \ref{simulations}b), although 
the fitting coefficients obtained numerically
do not correspond to those obtained experimentally. Moreover, the
monodisperse system shows similar results as long as the system is at
low osmotic pressure. Our results indicate that the significant
feature is not the detail of the system, but its proximity to the
jamming transition.

\section{Discussion and Conclusions}

We have presented experimental data showing the force distribution in
three dimensions of a lightly compressed emulsion, close to the
jamming transition. These data show an exponential distribution of
interdroplet forces $P(f)$ at large $f$. At low $f$, a peak in the
distribution function is observed. We have fitted the experimental
data with a function of the form $P(f)\propto
{f}^{0.9}e^{-1.9f/\bar{f}}$, suggested by the simple theoretical model
proposed for such a system. In addition we have carried out
simulations to determine the effect of polydispersity and osmotic
pressure on the force distribution function and these results are in
good agreement with the experimental data. They indicate that the
$P(f)$ is not sensitive to polydispersity, however the exponential
decay only fits the data well in distributions close to 'jamming', at
low confining pressures.

The theoretical model predicts a general distribution of the form
$P(f)\propto f^n e^{-(n+1)f/\bar{f}}$, where the power law coefficient
$n$ is determined by the packing geometry of the system. It is too
crude a model to account for the complexity of the emulsion system,
and it is therefore inappropriate to draw conclusions from direct
comparisons of the coefficients obtained from theory with those
arising from experimental and simulation data. Nevertheless, the
agreement in the functional form for all three methods is an important
result.  Curiously, we observe that the fitting coefficients agree for
the experimental data, the 2D theoretical model and the
quasi-monodisperse emulsion system at a comparable pressure to the
experiment.

In the future we hope to determine the mechanism by which the contact
patches between droplets exhibit enhanced fluorescence and also to use
the experimental data to test the validity of various force models for
compressible droplets.  In the experimental section we used the
Princen model to obtain the interdroplet force from the contact area
between particles. However in principle, we should be able to extract
the force law from the data.  This study gives supporting evidence to
the universality of the concept of jamming and provides a very
reliable experimental way of investigating microstructural elements
within the bulk of any refractive index matched, closely packed system
of an appropriate size.

We thank D. Bruji\'c, D. Grinev, 
J. Bibette, M. Shattuck, A. Tolley, J. Melrose, and R. Blumenfeld 
 for inspirational discussions. We are greatfull to the EPSRC and the Petroleum Research Fund for support of this work.

\newpage

\bibliographystyle{unsrt}

\begin{thebibliography}{}

\bibitem{Jamming}
A. Liu and S. R. Nagel, (eds.), {\it Jamming and Rheology:
Constrained Dynamics on Microscopic Scales}, Taylor \& Francis, London, 2001.

\bibitem{sam-zero}
S. F. Edwards and D. Grinev, {\it 
Advances in Complex Systems}, 2001, {\bf 4},  1.

\bibitem{Cipelletti2001}
V. Trappe, V. Prasad, L. Cipelletti, P. N. Segre and D. A. Weitz, {\it Nature}, 2001, {\bf 411}, 772.

\bibitem{Cates1999}
M. E. Cates, J. P. Wittmer, J. -P. Bouchaud and P. Claudin, {\it Chaos}, 1999,
 {\bf 9}, 511.

\bibitem{dantu} P. Dantu, {\it  G\'eotechnique}, 1968, {\bf 18},  50.

\bibitem{Liu1995}
C. H. Liu, S. R. Nagel, D. A. Schechter, S. N. Coppersmith, S. Majumdar, O.
Narayan, and T. A. Witten, {\it Science}, 1995,  {\bf 269},  513.

\bibitem{Radjai1996}
F. Radjai, M. Jean, J. Moreau, and S. Roux, {\it Phys. Rev. Lett.}, 1996, 
{\bf 77}, 274.

\bibitem{Thornton}
C. Thornton, {\it KONA Powder and Particle}, 1997, {\bf 15}, 81.

\bibitem{Makse2000}
H. A. Makse, D. L. Johnson, and L. M. Schwartz, {\it Phys. Rev. Lett.}, 2000,
 {\bf 84}, 4160.

\bibitem{Antony2001}
S. J. Antony, {\it Phys. Rev. E}, 2001,  {\bf 63}, 011302.


\bibitem{Ohern00}
C. S. O'Hern, S. A. Langer, A. J. Liu, and S. R. Nagel,
{\it Phys. Rev. Lett.}, 2001, {\bf 86}, 111.

 \bibitem{Coppersmith95}
S. N. Coppersmith, C.-H. Liu, S. Majumdar, O. Narayan and T. A. Witten,
{\it Phys. Rev. E}, 1995, {\bf 53}, 4673.

\bibitem{Mueth1998}
D. M. Mueth, H. M. Jaeger and S. R. Nagel, {\it Phys. Rev. E}, 1998,
{\bf 57},  3164.

\bibitem{Lovoll1999}
G. Lovoll, K. N. Maloy, E. G. Flekkoy, {\it Phys. Rev. E}, 1999,
{\bf 57}, 5872.

\bibitem{Blair2001}
D. L. Blair, N. W. Mueggenburg, A. H. Marshall, H. M. Jaeger, and S. R. Nagel,
{\it Phys. Rev. E}, 2001, {\bf 63}, 041304.

\bibitem{Sexton1999}
M. G. Sexton, J. E. S. Socolar, and D. G. Schaeffer, {\it
Phys. Rev. E}, 1999, {\bf 60},
1999.

\bibitem{Princen1983}
H. M. Princen, {\it J. Colloid Interface Sci.}, 1983, {\bf 91}, 160.

\bibitem{Bouchaud}
J.-P. Bouchaud, P. Claudin, D. Levine, and M. Otto, {\it Eur. J. Phys. E},
2001, {\bf 4}, 451.

\bibitem{faraday}
J. Bruji\'c, S. F. Edwards, D. V. Grinev, I.
Hopkinson, D. Bruji\'c, and H. A. Makse, submitted to {\it Faraday Disc.} 
(2002).

 \bibitem{Mason1995}
T. G. Mason, J. Bibette, and D. A. Weitz, {\it Phys. Rev. Lett.}, 1999,
 {\bf 75}, 2051.

\bibitem {Perrin2000}
T. G. Mason, and J. Bibette, {\it Langmuir}, 1997, 13, 4600.

\bibitem{Parker}
J. R. Parker, in {\it Algorithms for Image Processing and Computer Vision},
Wiley, New York, Chichester, 1997.

\bibitem{Witten1993}
D. C. Morse and T. A. Witten, {\it Europhys. Lett.}, 1993, {\bf 22}, 549.

\bibitem{Lacasse1995}
M-D. Lacasse, G. S. Grest, D. Levine, T. G. Mason, and D. A. Weitz, {\it Phys.
Rev. Lett.}, 1996, {\bf 76}, 3448.

\bibitem{Z}
S. F. Edwards and D. V. Grinev, {\it  Phys. Rev. Lett.}, 1999,  {\bf 82}, 5397.

\bibitem{cundall}
P. A. Cundall and O. D. L. Strack, {\it G\'eotechnique}, 1979,
{\bf 29}, 47.

\end{thebibliography}

\newpage

\begin{figure}
\centering
\caption{[See http://lisgi1.engr.ccny.cuny.edu/$\sim$makse/edwards/emulsions.html for this figure] 
Plot of the interdroplet forces
inside the packing of droplets. We plot only the forces larger than the
average for better visualisation.
Each rod joining the centers of two droplets in contact
represents a force. The thickness and the colour of the
rod is proportional to the magnitude of the force,
as obtained from the area of contact using Eq. (\protect\ref{force}).
}
\label{chains}
\end{figure}

\end{document}